\documentclass[]{aa}  
\usepackage{natbib}
 \usepackage{color}
     \usepackage{xcolor}
\bibpunct{(}{)}{;}{a}{}{,}
\newcommand{\rw}[1]{\textcolor{black}{#1}}
\newcommand{\mm}[1]{\textcolor{black}{#1}}
\newcommand{\ym}[1]{\textcolor{black}{#1}}
\newcommand{\nk}[1]{\textcolor{black}{#1}}

\usepackage{graphicx}

\begin{document}

   \title{H$_2$S and SO$_2$ detectability in Hot Jupiters}

   \subtitle{Sulfur species as indicator of metallicity and C/O ratio}

   \author{J. Polman
          \inst{1}
          \and
          L.B.F.M. Waters
          \inst{1,2}
          \and
          M. Min
          \inst{2}
          \and
          Y. Miguel
          \inst{3,2}
          \and
          N. Khorshid
          \inst{4}
          }

   \institute{Department of Astrophysics/IMAPP, Radboud University Nijmegen, PO Box 9010, NL-6500 GL Nijmegen, the Netherlands
         \and
        SRON Netherlands Institute for Space Research, Niels Bohrweg 4, NL-2333 CA Leiden, the Netherlands
        \and
        Sterrewacht leiden, University of Leiden,Niels Bohrweg 2, NL-2333 CA Leiden, The Netherlands
        \and
        Anton Pannekoek Institute for Astronomy, University of Amsterdam, Science Park 904, NL-1098 XH Amsterdam, the Netherlands
             }

   \date{Received ---; accepted ---}

\abstract
{The high cosmic abundance and the intermediate volatility and chemical properties of sulfur allow the use of sulfur-bearing species as a tracer of the chemical processes in the atmospheres of hot Jupiter exoplanets. Nevertheless, despite its properties and relevance as a tracer of the giant planets' formation history, little attention has been paid to this species in the context of hot Jupiter atmospheres.}
{In this paper, we provide an overview of the abundances of sulfur-bearing species in hot Jupiter atmospheres under different conditions and explore their observability.}
{We use the photochemical kinetics code VULCAN to model hot Jupiter atmospheric disequilibrium chemistry. Transmission spectra for these atmospheres are created using the modelling framework ARCiS. We vary model parameters such as the diffusion coefficient K$_{zz}$, and we study the importance of photochemistry on the resulting mixing ratios. Furthermore, we vary the chemical composition of the atmosphere by increasing the metallicity from solar to ~10 times solar. We also explore different C/O ratios.}
{We find that H$_2$S and SO$_2$ are the best candidates for detection between 1 and 10 $\mu$m, using a spectral resolution that is representative of the instruments on board the James Webb Space Telescope (JWST). H$_2$S is easiest to detect at an equilibrium temperature of $\sim$1500 K and C/O ratios between 0.7 and 0.9, with the ideal value increasing slightly for increasing metallicity. SO$_2$ is most likely to be detected at an equilibrium temperature of $\sim$1000 K at low C/O ratios and high metallicities. Nevertheless, among these two molecules, we expect SO$_2$ detection to be more common, as it is detectable in scenarios more favoured by formation models.}
{We conclude that H$_2$S and SO$_2$ will most likely be detected in the coming years with the JWST and that the detection of these species will provide information on atmospheric processes and planet formation scenarios.}

\maketitle
%

\section{Introduction}

Observations with Hubble, Spitzer and ground based telescopes have revealed a wide variety of atmospheric properties \citep{madhusudhan2019} and the more observations we have, the stronger the diversity observed in exoplanet atmospheres. The challenge is to understand this wide diversity, and place it in the context of planet formation scenarios \citep{SimAb2022,Turrini2021}. In particular, abundant species such as C-, N-, O- and S- bearing species can elucidate both the processes that occur in the atmosphere and at the same time provide clues on the formation history of the planet.

Sulfur is an important chemical element in this context. This is because of its cosmic abundance, its ability to bind to carbon and oxygen, its role in atmospheric chemistry, and its intermediate volatility when compared to highly refractory rock forming elements such as Fe, Mg and Si, and volatile species such as C and N. Recent planet formation models suggest that the combined use of the highly volatile N and the much more refractory S abundances in planetary atmospheres may be important to break degeneracies in planet formation scenarios when using only C and O \citep{Turrini2021}. Moreover, sulfur may behave refractory and could be used as a proxy for metallicity \citep{kama2019,Turrini2021}.

Atmospheric models of exoplanets have mainly focused on hydrogen, carbon, oxygen and nitrogen because they are the most abundant elements in the sun when excluding the non-reactive noble gases. This is due to the complexity of these models and the difficulty of obtaining accurate reaction rates. Nevertheless, efforts have been made to study sulfur chemistry \citep{Zahnle2009, Wang2017}, and sulfur networks are included in codes such as LEVI \citep{Hobbs2021} and VULCAN \citep{VULCAN2017,VULCAN2021}. Both models find that sulfur can play a significant role in the overall atmospheric chemistry by offering a faster pathway for liberating oxygen from H$_2$O.

Motivated by these efforts, in this paper we study the main sulfur-bearing species in hot Jupiter atmospheres and their potential detectability. We explore different potential scenarios such as different metallicities and C/O ratios, exploring the best conditions to detect these molecules and helping in the interpretation of future JWST data. A better quantitative understanding of the main sulfur reservoirs in molecular clouds, and planet forming disks is building up from ALMA and other millimetre wave telescopes \citep{laas2019,cazaux2022,legal2021,codella2021,cernicharo2021}, allowing in the future to put the sulfur abundance in exoplanetary atmospheres in context.

This paper is organised as follows: in Sect. 2 we describe the method we used to calculate the chemical composition of hot Jupiter atmospheres in the form of mixing ratios as a function of pressure, and the forward model we used to generate transmission spectra in the wavelength range of interest for observations with the James Webb Space Telescope (JWST). Section 3 describes the resulting atmospheric mixing ratios, and the emerging transmission spectra. We also vary some important parameters that affect the mixing ratios of molecular species of interest. In Sect. 4 we focus on H$_2$S and SO$_2$, the two most promising sulfur-bearing molecules with respect to  their detectability. In Sect. 5 we discuss our results, and Sect. 6 contains the conclusions of this study.

\section{Method}

\subsection{Disequilibrium chemistry calculations.}

We use the open-source photochemical kinetics code VULCAN \citep{VULCAN2017,VULCAN2021} to model hot Jupiter atmospheres. Within VULCAN we use a C-H-N-O-S chemical network, expanded from the C-H-N-O network of \citet{VULCAN2017}. The network consists of 87 species, approximately 500 forward chemical reactions, the associated reverse reactions, and 60 photodissociation branches. The included sulfur species are S, S$_{2}$, S$_{3}$, S$_{4}$, S$_{8}$, SH, HSO, H$_{2}$S, SO, CS, COS, CS2, NS, HS$_{2}$, SO$_{2}$, HCS, S$_{2}$O, CH$_{3}$SH and CH$_{3}$S. For the planetary parameters we use standard values for HD 189733 b as shown in Table \ref{tab:paraVULCAN}. The stellar flux of HD 189733 is taken from \citet{Moses2011}. We use protosolar elemental abundances from \citet{Lodders2009} (hereafter solar abundances). For the diffusion coefficient we assume a constant value of $10^9$ cm$^2$/s. The effect of this choice and the overall dependence of the results on the value of the diffusion coefficient are discussed in Sect. \ref{Discussion}. 

We use 150 vertical atmospheric layers, resulting in roughly 6 layers per pressure scale height, as recommended for VULCAN to ensure visibly smooth profiles. For each layer the number densities of each species are computed based on the number densities at the preceding timestep, the reaction rates of each reaction, the stellar flux reaching the layer for photodissociation, and the diffusion coefficient for mixing between each layer. For the convergence parameters we use the recommended values $\delta$=0.01 and $\epsilon$=10$^{-4}$ s$^{-1}$ associated with the conditions $\Delta \hat{n}$\textless$\delta$ and $\frac{\Delta\hat{n}}{\Delta t}$\textless$\epsilon$, where $\Delta\hat{n}\equiv\frac{\mid n_{i,j,k} - n_{i,j,k'}\mid}{n_{i,j,k}}$ and $\Delta t\equiv t_k - t_{k'}$, with i denoting the species, j the layer and k the timestep. The convergence criteria ensure consistent results and are the same as those recommended by \citet{VULCAN2017}. We found that the relative tolerance for adjusting the step size had to be changed often to ensure convergence without large elemental losses. Simulations were redone with different values for this tolerance if elemental losses or gains of more than one percent occurred. A summary, alongside other parameters used in VULCAN, is shown in Table \ref{tab:paraVULCAN}.

\begin{figure}
    \resizebox{\hsize}{!}{\includegraphics{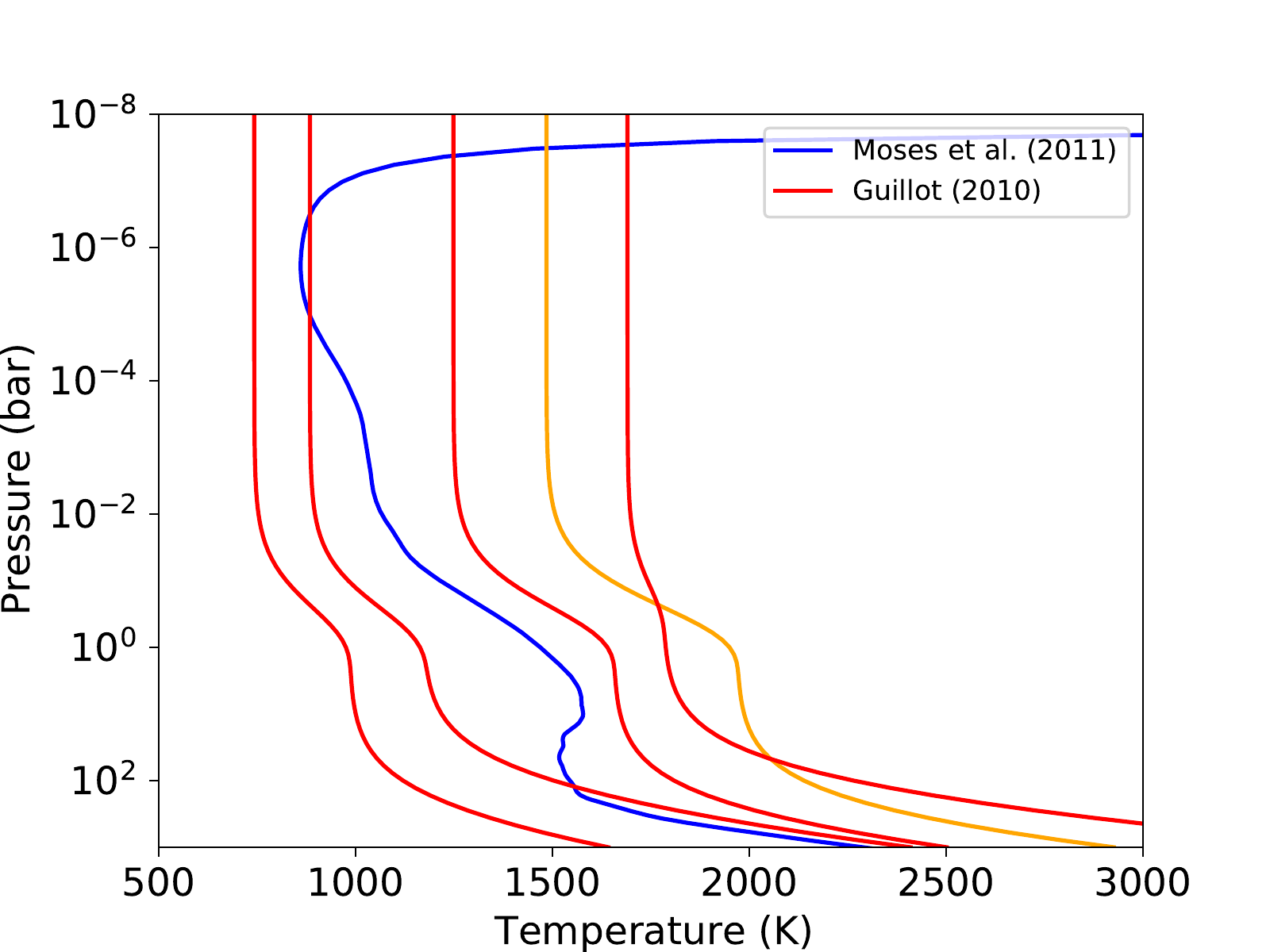}}
        \caption{TP-profiles used to understand temperature dependencies of sulfur-bearing species. When analysing the C/O ratio and metallicity dependence of H$_2$S we will mainly consider the orange profile. Similarly for SO$_2$ we mainly consider the blue TP-profile.}
            \label{TPprofiles}
    \end{figure}

We will mainly focus on two different temperature-pressure (TP) profiles, a TP-profile from \citet{Moses2011} for HD 189733 b, and another parametrized using Eq.~(29) from \citet{Guillot2010}. The used parameters are the mean thermal opacity $\kappa_{th}$, the ratio between the mean visible opacity and the mean thermal opacity $\gamma$, the flux parameter f and the intrinsic temperature T$_{int}$. The TP-profiles used are shown in Fig. \ref{TPprofiles}, with the values for the parameters given in Table \ref{table_Guillot}. The reasons for the choices made here, and the dependence of the results on these choices, are discussed in Sect. \ref{Discussion}.

\begin{table}[]
\caption{Standard parameters used within VULCAN.}
\label{tab:paraVULCAN}
\centering
\begin{tabular}{|c|c|}
\hline
He/H                      & 0.0975                                                \\
C/H                       & 2.776$\cdot10^{-4}$   \\
N/H                       & 8.185$\cdot10^{-5}$   \\
O/H                       & 6.062$\cdot10^{-4}$   \\
S/H                       & 1.626$\cdot10^{-5}$   \\ \hline
R$_\star$   & 0.805 R$_\odot$                         \\
R$_p$                      & 1.138 R$_J$                                            \\
Orbital radius            & 0.03142 au                                          \\
Zenith angle              & 48\textdegree                            \\
Eddington factor     & 0.5                                                   \\
Number of vertical layers & 150                                                   \\
P$_{bottom}$             & 10$^3$ bar      \\
P$_{top}$                & 10$^{-8}$ bar \\
K$_{zz}$                 & 10$^9$ cm$^2$/s                                \\
Surface gravity           & 2140 cm/s$^2$                          \\ 
$\delta$ & 0.01 \\
$\epsilon$ & 10$^{-4}$ s$^{-1}$ \\\hline
\end{tabular}
\end{table}

\begin{table*}
\caption{The values for the parameters used to create the temperature-pressure profiles using the method of \citet{Guillot2010}. Temperature-pressure profiles are ordered from coldest to hottest at the top of the atmosphere.}
\label{table_Guillot}
\centering
\begin{tabular}{|c|c|c|c|c|c|}
\hline
$\kappa_{th}$ (cm$^{2}$g$^{-1}$) & 1.40$\cdot$10$^{-2}$ & 1.40$\cdot$10$^{-2}$ & 1.40$\cdot$10$^{-2}$ & 1.40$\cdot$10$^{-2}$  & 1.40$\cdot$10$^{-2}$  \\
$\gamma$           & 0.3 & 0.3    & 0.3   & 0.3    & 0.8    \\
f               & 0.05   & 0.1    & 0.4    & 0.8   & 1         \\
T$_{int}$ (K)         & 200       & 300       & 300       & 350       & 450       \\ \hline
\end{tabular}
\end{table*}

\subsection{Modeling the synthetic spectra.} \label{sec:ARCiS}

Having obtained mixing ratios using VULCAN, we use the ARCiS modelling framework \citep{ARCiS2019,ARCiS2020} to create transmission spectra based on these mixing ratios. Within ARCiS we again use standard values for the planetary parameters of HD~189733 b. ARCiS uses correlated k-tables to create transmission spectra at low spectral resolutions (R$\leq$1000). We focus on a wavelength range of 1 to 10 $\mu$m. We have opacity data for 26 of the 87 species incorporated in the VULCAN C-H-N-O-S network. \mm{The correlated-k data were all taken from the ExoMolOP database \citep{2021A&A...646A..21C}}. This data includes the most abundant and relevant species, with the exception of SO. We discuss the effect of not being able to include SO in Sect. \ref{Discussion}. 

To determine the detectability of a species, we consider the difference between transmission spectra including and excluding the relevant species. For each species we first analyse the entire wavelength range to determine at which wavelength the difference between the spectra is the largest. Simulations with different parameters are compared for the specific wavelength bin at which the difference is generally the largest, 3.78 $\mu$m for H$_2$S and 7.26 $\mu$m for SO$_2$. The difference between the spectrum with the species and without is expressed in parts per million of the stellar flux. We will refer to this difference as the detectability. We acknowledge that this does not truly reflect how likely it is to retrieve species from transmission spectra, since it ignores other possible sources of high opacity at this wavelength and the effect of the opacity at other wavelengths, as well as other reasons. We leave in depth analysis using retrieval methods for specific planets to a future study. 

\section{General results}

    \begin{figure}
    \resizebox{\hsize}{!}{\includegraphics{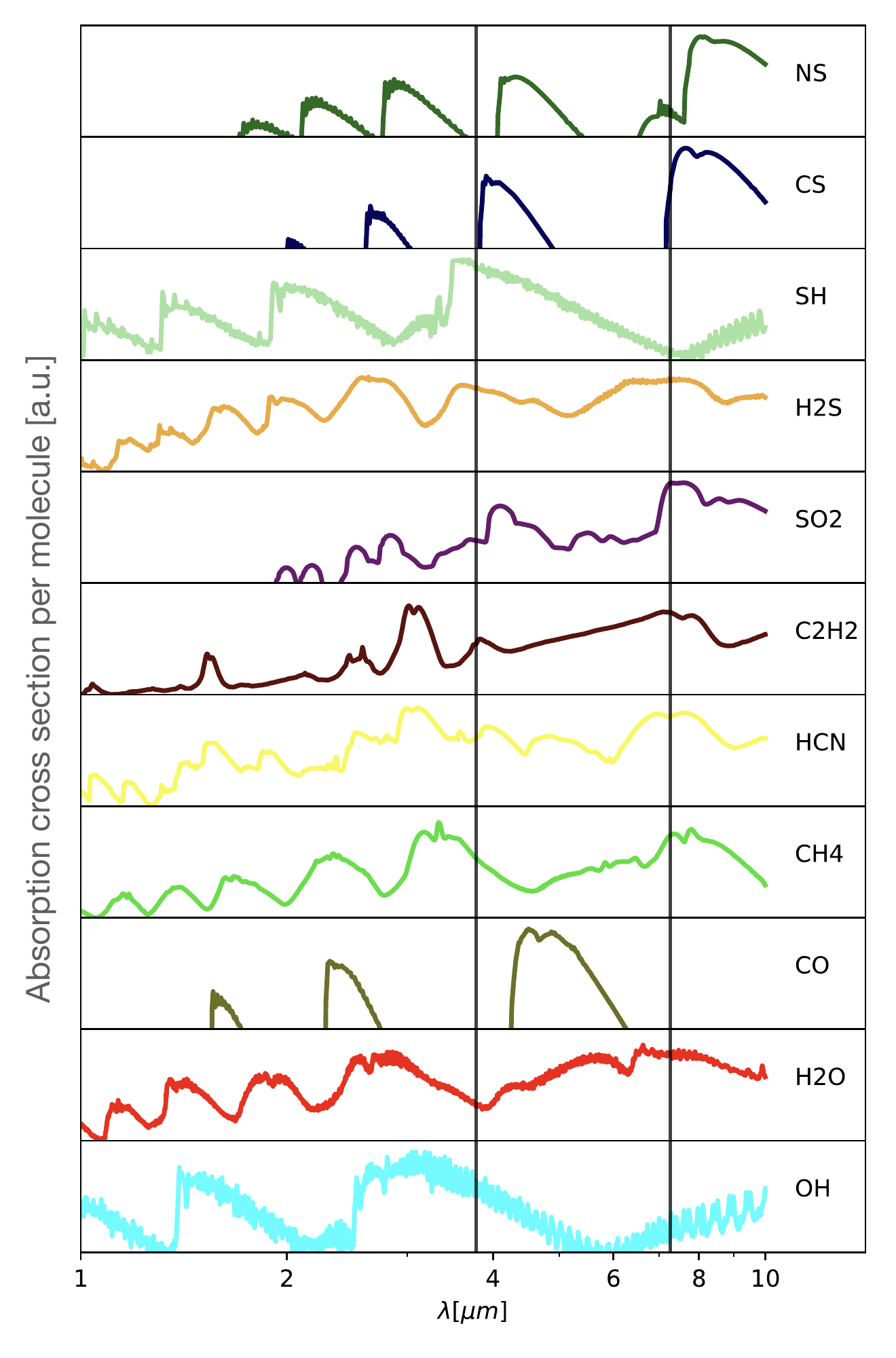}}
        \caption{Opacities of relevant species for a pressure of 1 bar and a temperature of 1000 K, averaged over the values of the correlated k-tables. The black lines indicate 3.78 and 7.26 $\mu$m, the wavelengths at which we analyse the detectability of H$_2$S and SO$_2$ respectively.}
        \label{Opacities}
    \end{figure}
    
\begin{figure*}
     \includegraphics[width=17cm]{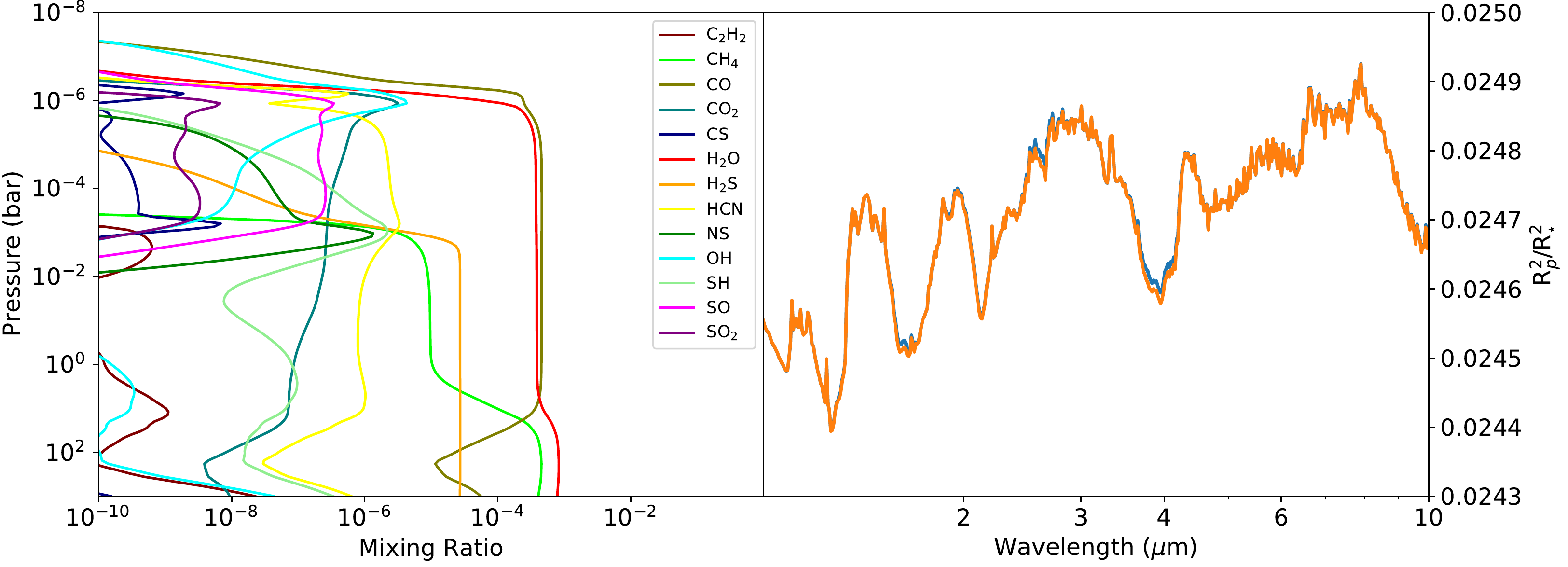}
        \caption{Mixing ratios for the most relevant species (left) and transmission spectrum (right) with and without H$_2$S. Model parameters are given in Table \ref{tab:paraVULCAN}}
            \label{Example_mr_and_trans}
    \end{figure*}

\rw{We start out by discussing the resulting transmission spectrum for our standard model (parameters given in Table \ref{tab:paraVULCAN}), and assuming solar abundances for the elements.} Figure \ref{Example_mr_and_trans} shows mixing ratios of the most relevant species for this study for solar abundances and the temperature profile of HD 189733 b. At the highest pressures most carbon is stored in CH$_4$. This changes around 10$^1$ bar, with most carbon being stored in CO at lower pressures, with a large part of the remaining oxygen being stored in H$_2$O, which shows similar abundances to CO, due to the C/O ratio being close to 0.5. H$_2$S is by far the most abundant sulfur-bearing species. This result is consistent across all tested metallicities, C/O ratios and temperatures. 

Figure \ref{Example_mr_and_trans} also shows the effect H$_2$S has on the transmission spectrum, by comparing transmission spectra with and without considering H$_2$S. This effect is quite small, peaking at 14 ppm. The difference is small, because H$_2$S is obscured by other species, mainly H$_2$O and HCN. The opacities of these species are shown in simplified form in Fig. \ref{Opacities}. This illustrates the two main requirements for being able to detect a species, i.e. (i) a high enough mixing ratio to have an effect on the transmission spectrum at a specific wavelength, and (ii) not having other species competing at this wavelength. Species located higher in the atmosphere are less likely to be obscured, since these do not compete with those lower in the atmosphere. Species located lower in the atmosphere, such as H$_2$S in this case, can still be detected as long as other species in the atmosphere do not compete at the relevant wavelength.

We now consider CS, H$_2$S, NS, SH and SO$_2$, the sulfur-bearing species for which we have opacity data. We analyse their detectability for six different TP-profiles, shown in Fig. \ref{TPprofiles}, and a wide range of C/O ratios and metallicities. 

\subsection{NS}
NS can become the main sulfur reservoir at pressure between 10$^{-5}$ and 10$^{-3}$ bar for colder planets with an equilibrium temperature between 750 K and 1000 K, although this does require a diffusion coefficient of 10$^{11}$ cm$^2$/s. 750 K is the lowest temperature we considered, so we do not how it behaves at lower temperatures. This does not lead to NS being detectable, even with such strong diffusion, due to its low opacity and the opacity being highest between a wavelength of 8 and 9 $\mu$m. At this wavelength C$_2$H$_2$, NH$_3$, HCN and CH$_4$ also have a high opacity, leading to NS being obscured.

\subsection{CS}
Similarly, CS can also become the main sulfur reservoir, although this happens at lower pressures than for NS, between 10$^{-6}$ and 10$^{-4}$ bar. In this case for planets with an equilibrium temperature of about 900 K for solar elemental abundances. Again, just as for NS, this only happens for diffusion coefficients of at least 10$^{11}$ cm$^2$/s. For extremely carbon enriched atmospheres, CS thrives at higher temperatures, with it being the main sulfur reservoir between 10$^{-6}$ and 10$^{-1}$ bar. CS detection suffers the same problems as NS, with a slightly higher, but still low opacity, with the opacity being greatest near 8 $\mu$m. In the carbon enriched, high temperature case, CS is obscured at this wavelength by C$_2$H$_2$ and CH$_4$. 

\subsection{SH}
The mixing ratio of the third species we consider, SH, is very dependent on the mixing ratio of H$_2$S. It is only ever the main sulfur reservoir for a very specific pressure range, at pressures lower than the pressure at which H$_2$S becomes less abundant, at around 10$^{-3}$ bar. For higher temperatures this happens slightly higher in the atmosphere. This is in agreement with previous studies into sulfur chemistry \citep{Zahnle2009,Wang2017,Hobbs2021,VULCAN2021}. The opacity of SH is highest just below 4 $\mu$m, similar to H$_2$S, with a similar value. Since H$_2$S is significantly more abundant, it will always be easier to detect at this wavelength. As SH shows no other strong opacity features between 1 and 10 $\mu$m, we shall not further consider it in our analysis.

\subsection{H$_2$S}
As mentioned before, H$_2$S is the main sulfur reservoir below 10$^{-3}$ bar, with nearly all sulfur being in the form of H$_2$S across all tested temperatures from an equilibrium temperature of 750 K to 1700 K. Just like for SH, this is in full agreement with previous studies \citep{Zahnle2009,Wang2017,Hobbs2021,VULCAN2021}. Since the abundance of H$_2$S is almost entirely dependent on the abundance of sulfur, and almost entirely temperature and C/O ratio independent, with its detectability largely determined by the abundance of other species that can obscure it higher up in the atmosphere.

\begin{figure*}
    \centering
        \includegraphics[width=17cm]{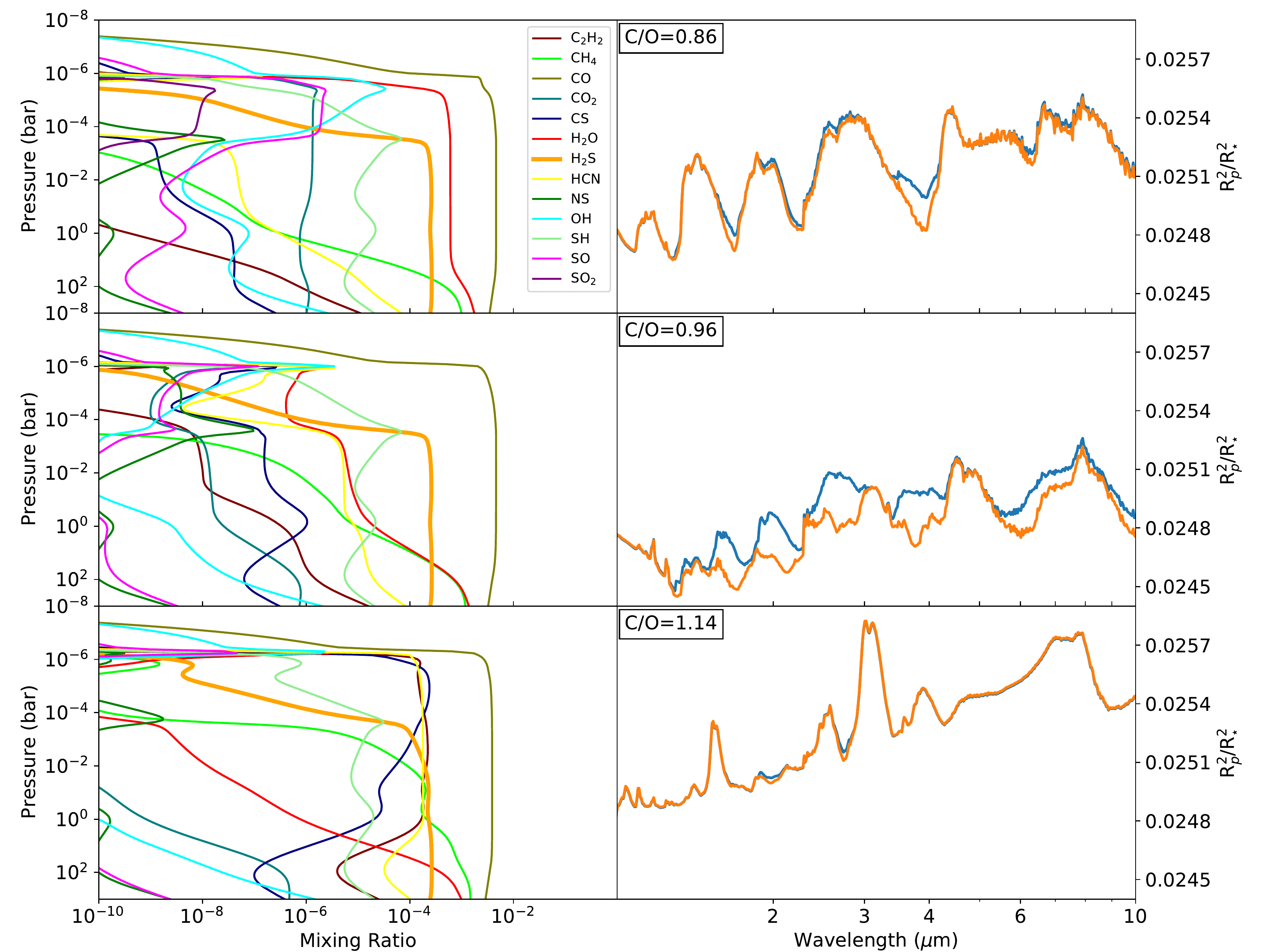}
            \caption{Illustration of the detectability of H$_2$S around C/O=0.9 for 10 times solar metallicity. From top to bottom C/O=0.86, 0.96, and 1.14. The spectra on the right show the transit spectrum including or excluding H$_2$S opacity in blue and orange respectively.}
            \label{H2S_mr_Example}
    \end{figure*}

H$_2$S has a high opacity for a wide range of wavelengths (Fig.~\ref{Opacities}), but the opacity near 3.8 $\mu$m turns out to be the most relevant, due to H$_2$S being obscured at most other wavelengths. For most temperatures and C/O ratios H$_2$S is largely obscured, but for equilibrium temperatures between 1250 K and 1700 K an interesting effect can be observed, which is shown in Fig. \ref{H2S_mr_Example}. For a C/O ratio of about 0.9, \rw{the chemistry of the atmosphere begins to switch from O--rich to C-rich, and this is reflected in the overall shape of the transmission spectrum. In particular, } we can very clearly see the effect of H$_2$S on the transmission spectrum. At this C/O ratio almost all carbon and oxygen is stored in CO, which has a very low opacity at the relevant wavelength of 3.8 $\mu$m. For a higher C/O ratio, C$_2$H$_2$ and HCN become more abundant, obscuring H$_2$S. Lower C/O ratios cause H$_2$O to become more abundant, which also obscures H$_2$S, although its opacity is lower than C$_2$H$_2$ and HCN near 3.8 $\mu$m. This effect is elaborated on in Sect. \ref{Results}. 

\subsection{SO$_2$}
Similar to CS and NS, SO$_2$ is abundant between 10$^{-6}$ and 10$^{-3}$ bar, although almost always less abundant than or just as abundant as SO, for which we do not have opacity data. SO$_2$ is most abundant at equilibrium temperatures between 900 K and 1000 K. Since SO$_2$ is expected to be located higher in the atmosphere than H$_2$S, it is less likely to be obscured, leading to its detectability being mostly dependent on its abundance alone. This is strengthened by the fact that the opacity of SO$_2$ is highest between 7 and 8 $\mu$m, a wavelength at which most species show low opacities. 

\begin{figure*}
    \centering
        \includegraphics[width=17cm]{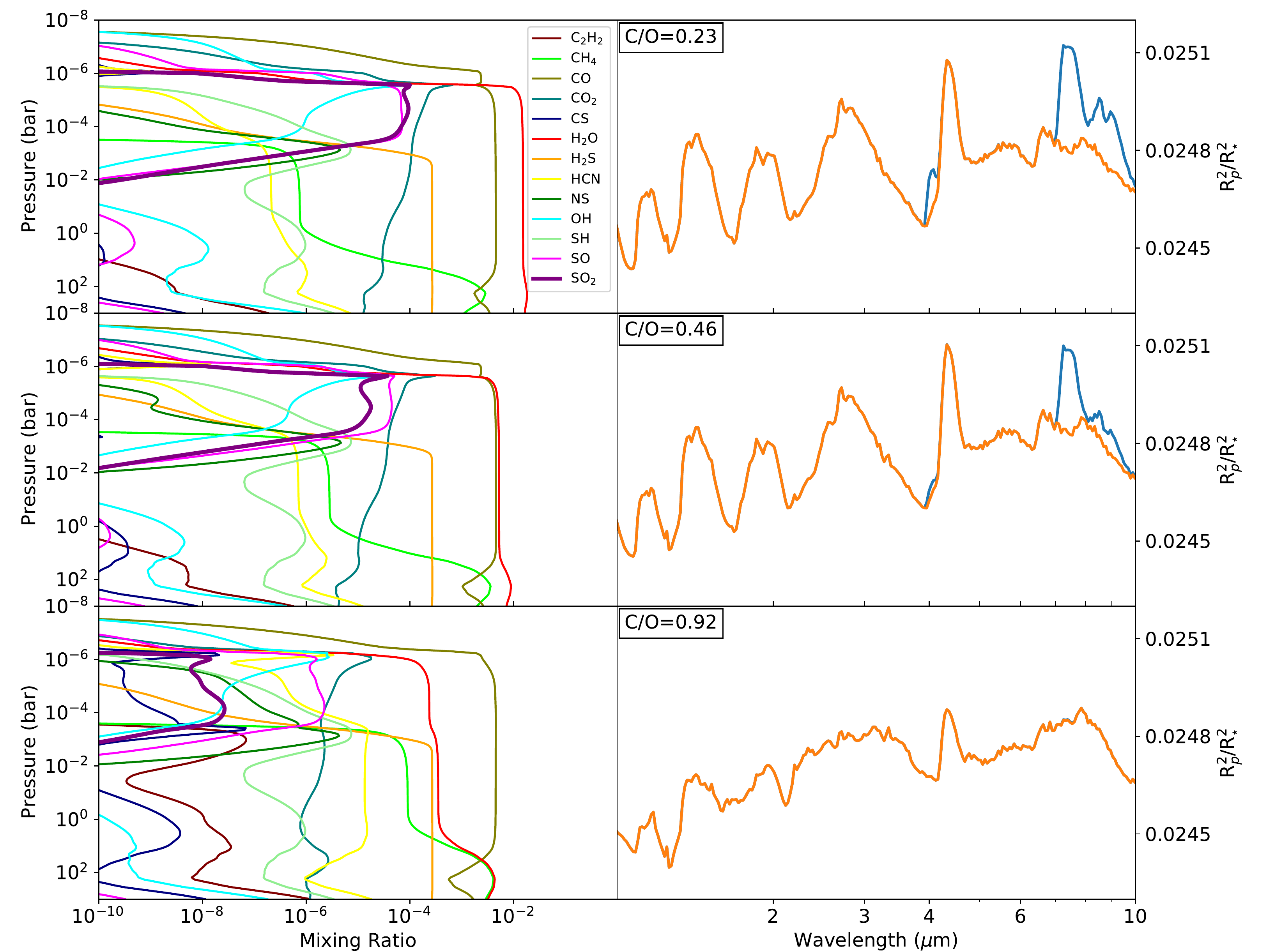}
            \caption{Increased detectability of SO$_2$ shown for 10 times solar metallicity for C/O ratios of 0.23, 0.46 and 0.92 (top to bottom).}
            \label{SO2_mr_Example}
    \end{figure*}

\begin{figure*}
    \centering
    \includegraphics[width=17cm]{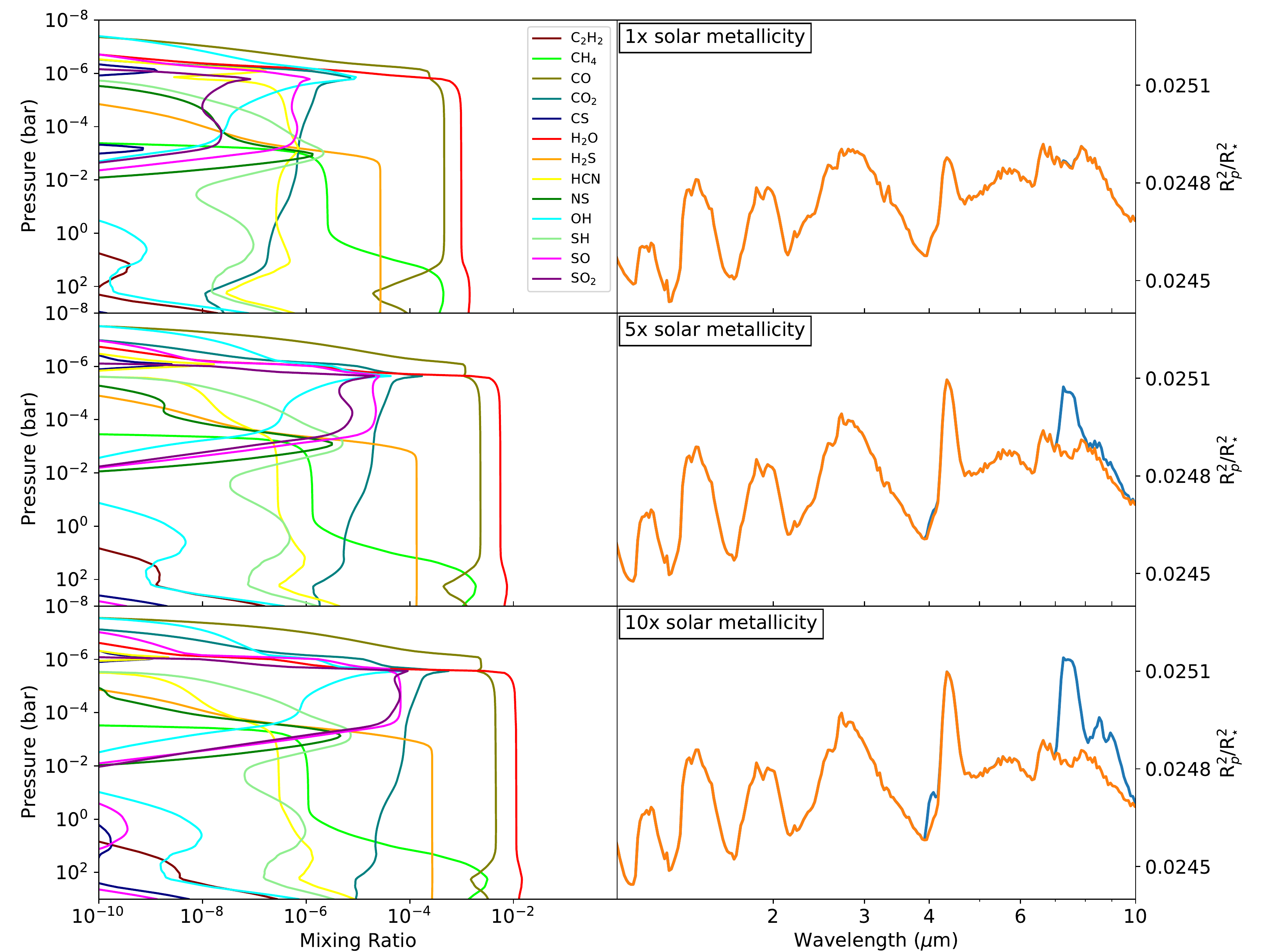}
        \caption{Increased detectability of SO$_2$ shown for a C/O ratio of 0.29, metallicity 1,5 and 10 times solar.}
            \label{SO2_mr_met_Example}
    \end{figure*}

We again observe an interesting effect. Whereas the abundance of H$_2$S is extremely independent of the abundance of the elements besides sulfur, this is not the case for the SO$_2$ abundance.  Lower C/O ratios and higher metallicities both lead to an increase in the abundance of SO$_2$. For lower C/O ratios this can be explained by an increase in available oxygen, with the rest being stored in CO and CO$_2$. This effect is shown in Fig. \ref{SO2_mr_Example}. Similar to how more available oxygen leads to a higher SO$_2$ abundance, a higher metallicity has the same effect, providing more sulfur and oxygen which leads to a higher abundance of SO$_2$. Naively one might expect the abundance of SO$_2$ to be proportional to the elemental abundance of sulfur and oxygen, but it turns out the effect is significantly stronger. This is shown in Fig. \ref{SO2_mr_met_Example}. 

\begin{figure}
    \resizebox{\hsize}{!}{\includegraphics{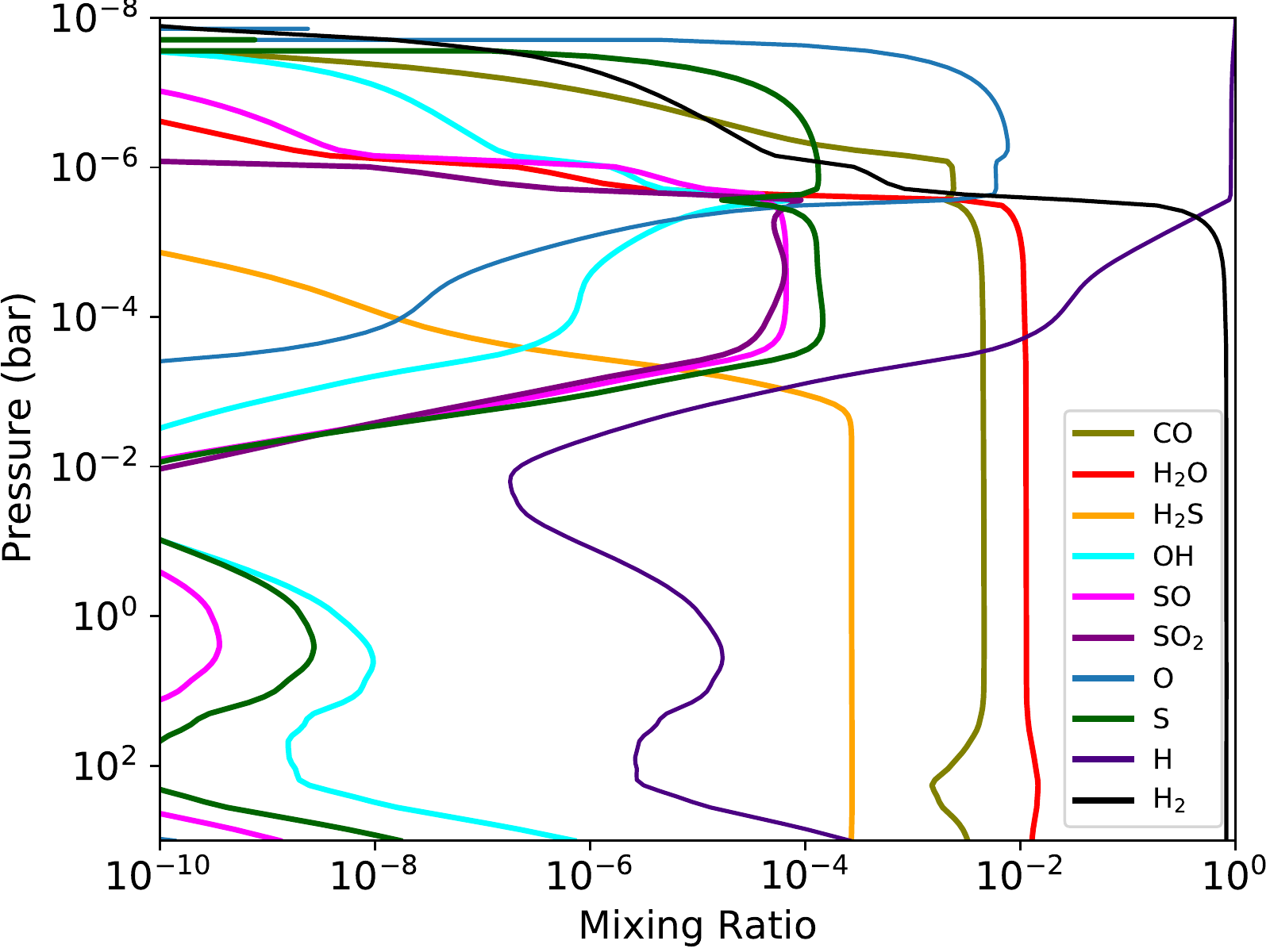}}
        \caption{Mixing ratios of the main species relevant for SO$_2$ creation for a C/O ratio of 0.29 and 10 times solar metallicity.}
            \label{SO2_species_in_depth}
    \end{figure}

This effect is best understood by analysing the relevant reactions for SO$_2$. At the relevant pressures and mixing ratios, the reaction SO+OH\textrightarrow SO$_2$+H dominates by several orders when ignoring photodissociation. Photodissociation dominates the loss of SO$_2$. The balance between this formation and destruction determines the steady state abundance. At these low pressures the previously discussed species no longer play a significant role, since H$_2$S and SH do no exist with large mixing ratio. The only important species are S, S$_2$ and the SO and SO$_2$ discussed here. SO is mainly created through OH+S\textrightarrow SO+H and through the photodissociation of SO$_2$. SO is only lost through creating SO$_2$ and through photodissociation. OH, vital for the creation of SO and SO$_2$, is primarily formed through the photodissociation of H$_2$O and through O+H$_2$\textrightarrow OH+H. This explains both the metallicity and the C/O ratio dependence of the SO$_2$ abundance. A low C/O ratio leads to a large abundance of H$_2$O, providing more OH to produce SO and SO$_2$. High metallicities also lead to a larger H$_2$O abundance, while also increasing the availability of S, necessary to produce SO. In the figure we can see that the overall effect of the C/O ratio and metallicity is a lot stronger for SO$_2$ than for SO. The SO/SO$_2$ ratio, decreases from $\sim$50 at 10$^{-5}$ bar at solar metallicity, with increased oxygen to reach the C/O ratio of 0.29, to $\sim$1 for 10 times solar metallicity, with the same relative increase in oxygen. Figure \ref{SO2_species_in_depth} shows an example of the mixing ratios of all the relevant species in this process. At 10$^{-6}$ bar a peak in OH can be seen, which coincides with a small dip in the abundance of S and the photodissociation of H$_2$O. At lower pressures SO$_2$ cannot exist due to the stellar irradiation. At higher pressures, due to some photodissociation of H$_2$O still occurring and due to vertical mixing, SO$_2$ is also formed and mixed to higher pressures. The mixing ratio sharply decreases at 10$^{-3}$ bar due to almost all available sulfur forming H$_2$S, thus providing the upper pressure limit for SO$_2$.

\subsection{Summary of detectability}
Summarising, we find that of the sulfur-bearing molecular species in the atmospheres we have studied, and for which we have opacity data, H$_2$S and SO$_2$ are the most likely to be detected in low resolution  (R$\sim$100-1000) transmission spectra in the 1-10 $\mu$m wavelength range. In what follows, we will study these two molecules in more detail.

\section{In depth analysis H$_2$S and SO$_2$} \label{Results}
Having discussed why we consider H$_2$S and SO$_2$ the most interesting sulfur species for detection at low resolution at a wavelength of 1 to 10 $\mu$m, we continue to analyse the observed effects for these species in greater detail. The effect of the C/O ratio and metallicity on the detectability is first analysed for H$_2$S and then for SO$_2$.

\subsection{H$_2S$} \label{In_depth_H2S}
\begin{figure*}
    \sidecaption
        \includegraphics[width=12cm]{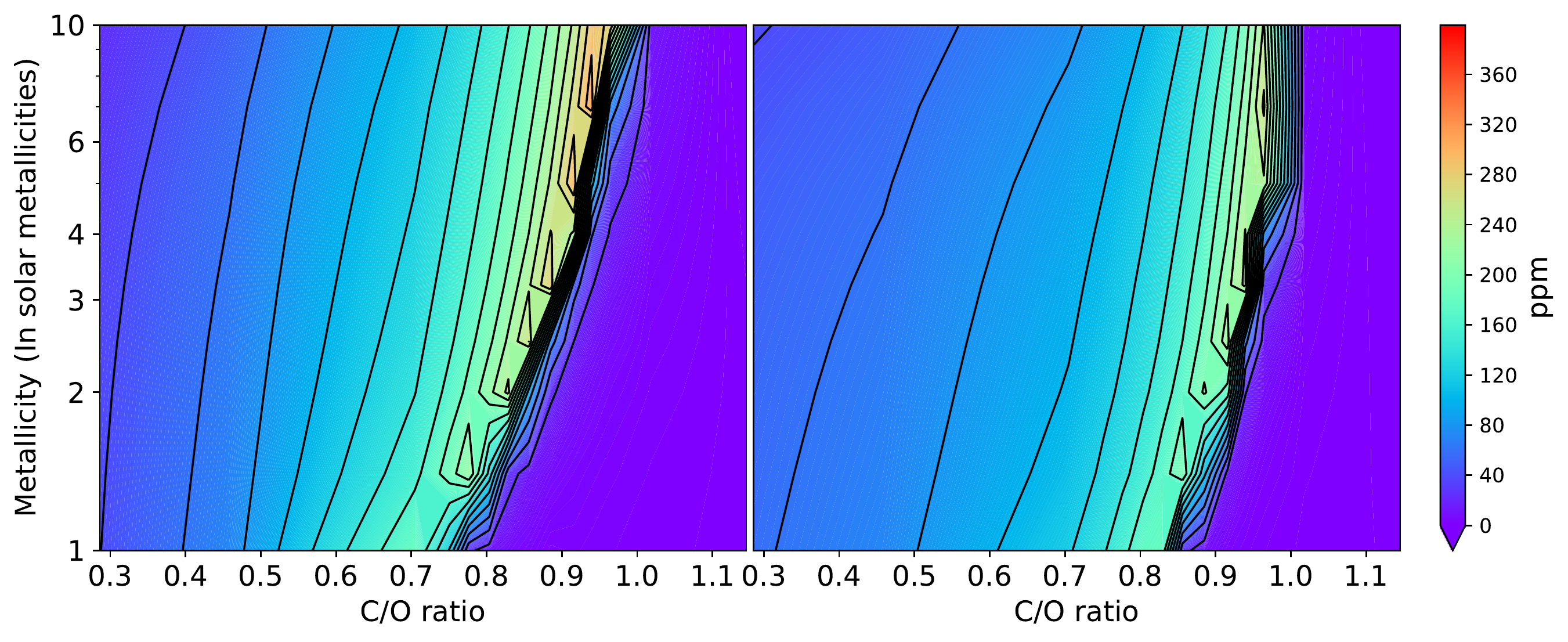}
        \caption{The difference between transmission spectra with and without H$_2$S for the wavelength bin centred on 3.78 $\mu$m as a function of metallicity and C/O ratio. This difference gives the detectability as described in Sect. \ref{sec:ARCiS}. We vary the C/O ratio by changing the the oxygen (left) and carbon (right) abundance.}
        \label{H2S_Contour_Example}
    \end{figure*}

Figure \ref{H2S_Contour_Example} shows the difference between transmission spectra including and excluding H$_2$S for the spectral bin centred on 3.78 $\mu$m for a spectral resolution of 200 for a wide range of metallicities and C/O ratios. As mentioned before, this value near 3.8 $\mu$m was chosen, since in most cases this shows the largest difference between the two spectra, while also clearly showcasing the effect.  Of course H$_2$S can also have a high detectability for different wavelengths, especially near C/O$\sim$0.9, as can be seen in Fig. \ref{H2S_mr_Example}. As Fig. \ref{H2S_Contour_Example} shows, the detectability increases for an increasing C/O ratio until it peaks and then quickly decreases. This is caused, by first the abundance H$_2$O decreasing, and after the peak the abundance of C$_2$H$_2$ and HCN increasing. What can also be seen is that the peak is slightly metallicity dependent, with the peak moving to higher C/O ratios for an increase in metallicity. The actual value of this peak is only slightly metallicity dependent, which can be explained via the behaviour of H$_2$S. Low in the atmosphere almost all sulfur is stored in H$_2$S, completely blocking all light at a wavelength of 3.8 $\mu$m, an increase in metallicity only has a small effect on the impact of H$_2$S on the transmission spectrum. 

The detectability peaks at a value of 314 ppm for the case that the oxygen abundance is varied. The change in C/O ratio is achieved by either changing the abundance of carbon or oxygen. Both achieve a very similar result, showing that the effect is mainly dependent on the C/O ratio, and not just the abundance of carbon or oxygen. Knowing this we will, as done in previous plots, from now on only show results achieved by varying the abundance of oxygen.

\subsection{SO$_2$}
\begin{figure*}
        \sidecaption
        \includegraphics[width=12cm]{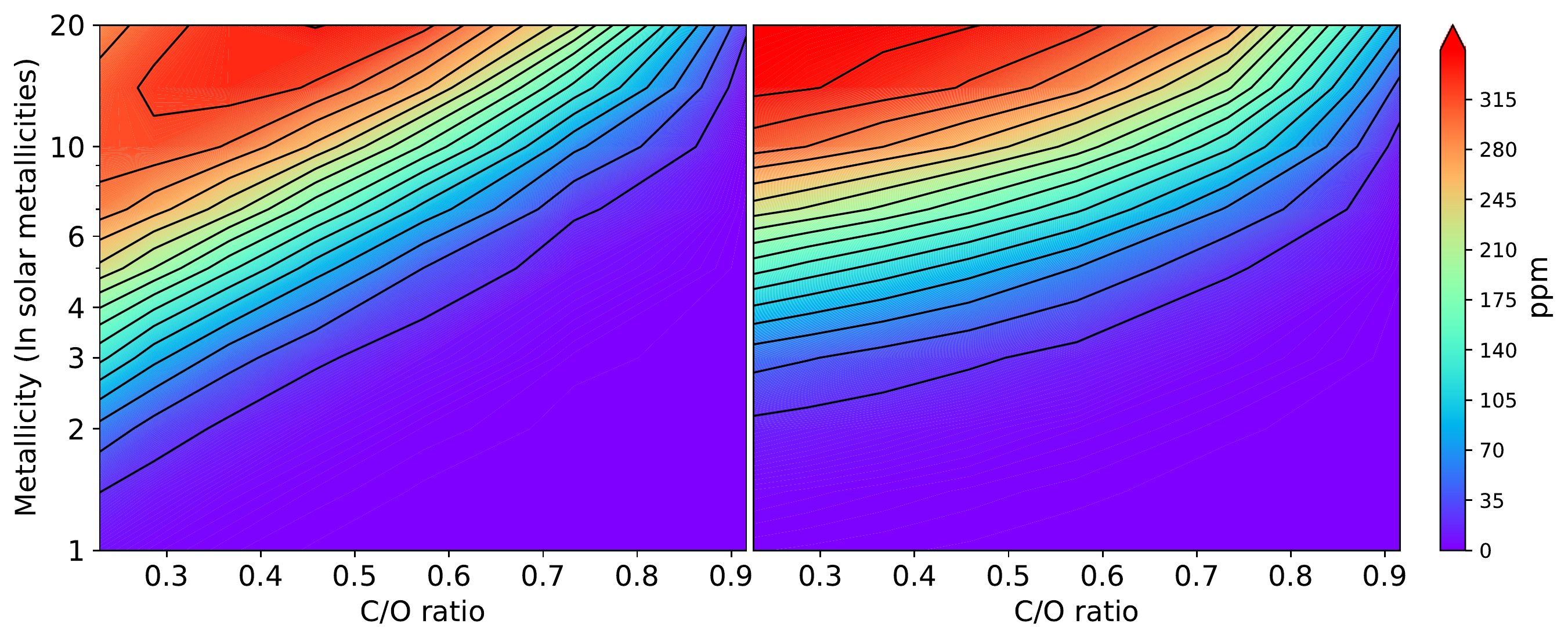}
        \caption{Same as Fig. \ref{H2S_Contour_Example} for SO2, using the 7.26 um spectral region.}
        \label{SO2_Contour_Example}
    \end{figure*}

Figure \ref{SO2_Contour_Example} shows the difference between transmission spectra including and excluding SO$_2$ for a bin centred on 7.26 $\mu$m for a spectral resolution of 100 for a wide range of metallicities and C/O ratios. This wavelength was chosen, because the difference between the transmission spectra is largest here for most simulations, while it only differs a few percent in the cases that it is not. This  shows the same effect described in the previous section. The overall detectability of SO$_2$ is higher for low C/O ratios and high metallicities. At C/O ratios near 1 the SO$_2$ abundance is never high enough for it to be detectable, since almost all oxygen will be stored in CO. Similarly, for solar metallicities, detecting SO$_2$ is not possible, with the effect on the transmission spectrum not exceeding 10 ppm. This is the result of what was described in the previous section, detailing the relation between the abundance of SO$_2$, the abundance of species reacting to form SO$_2$, and the effect metallicity has on these abundances. 

At the highest metallicities and lowest C/O ratios some irregularities can be observed that do not follow the expected pattern. These can be explained by the maximum effect of SO$_2$ on the transmission being reached, with all light at 7.26 $\mu$m in the relevant pressure range being blocked. Higher SO$_2$ abundances no longer have an effect on the transmission spectra when this is the case. At this point slight increases in the abundances of other species can have a small effect on the detectability, due to the way we derived this value, whereas normally these small changes would be unnoticeable due to the increase of the effect SO$_2$ has on the spectrum. The detectability peaks at a value of 341 ppm when varying the oxygen abundance. Similar to the situation for H$_2$S, we again observe little difference between the contour plots derived using either a variation of carbon or oxygen to achieve the desired C/O ratio, showing that the C/O ratio is more relevant than the abundance of oxygen or carbon individually. As done up until this point, and as decided on for H$_2$S, we will from now on only show results in which the C/O ratio was achieved by varying the abundance of oxygen.

\section{Discussion} \label{Discussion}

\subsection{Temperature dependence}
\begin{figure*}
        \sidecaption
        \includegraphics[width=12cm]{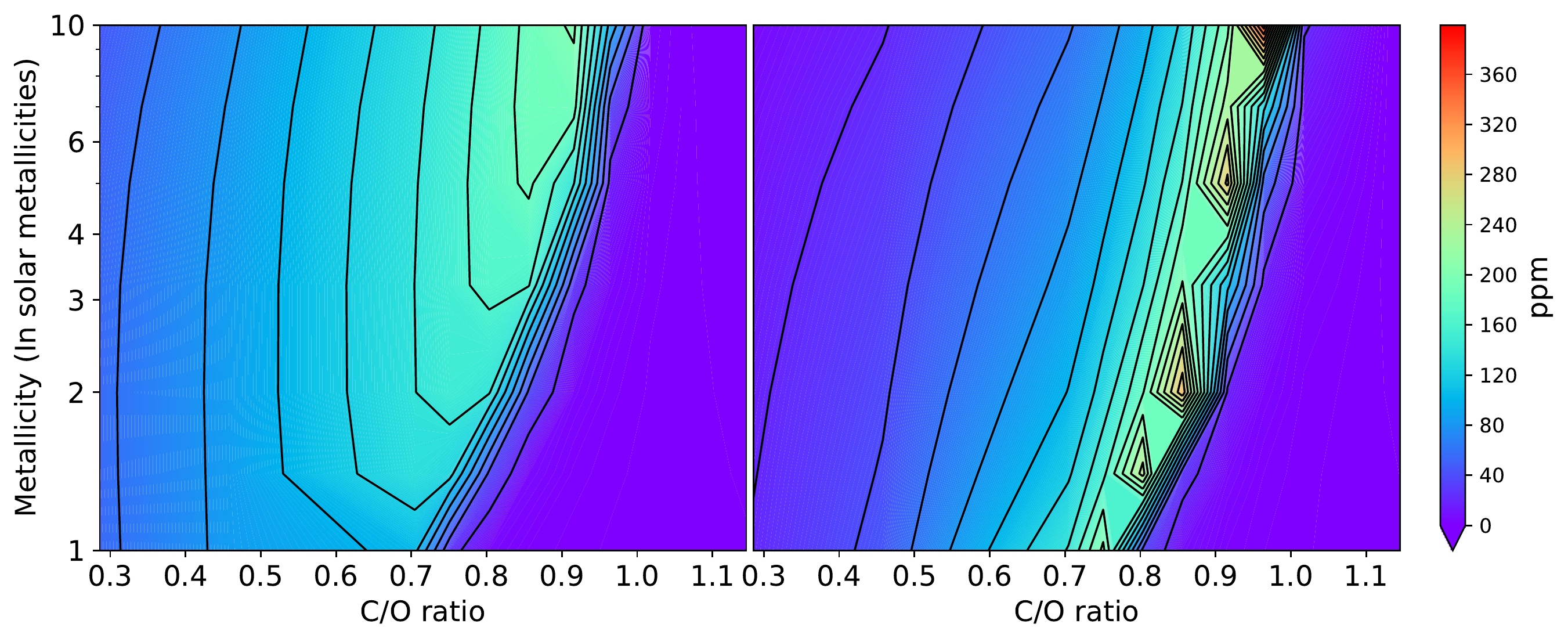}
        \caption{Contour plot of H$_2$S detectability similar to Fig. \ref{H2S_Contour_Example} for two TP-profiles changed to be 200 K colder (left) or hotter (right) in order for all pressures. In both cases the C/O ratio is varied by changing the oxygen abundance.}
        \label{H2S_temp_change}
    \end{figure*}

\begin{figure*}
        \sidecaption
        \includegraphics[width=12cm]{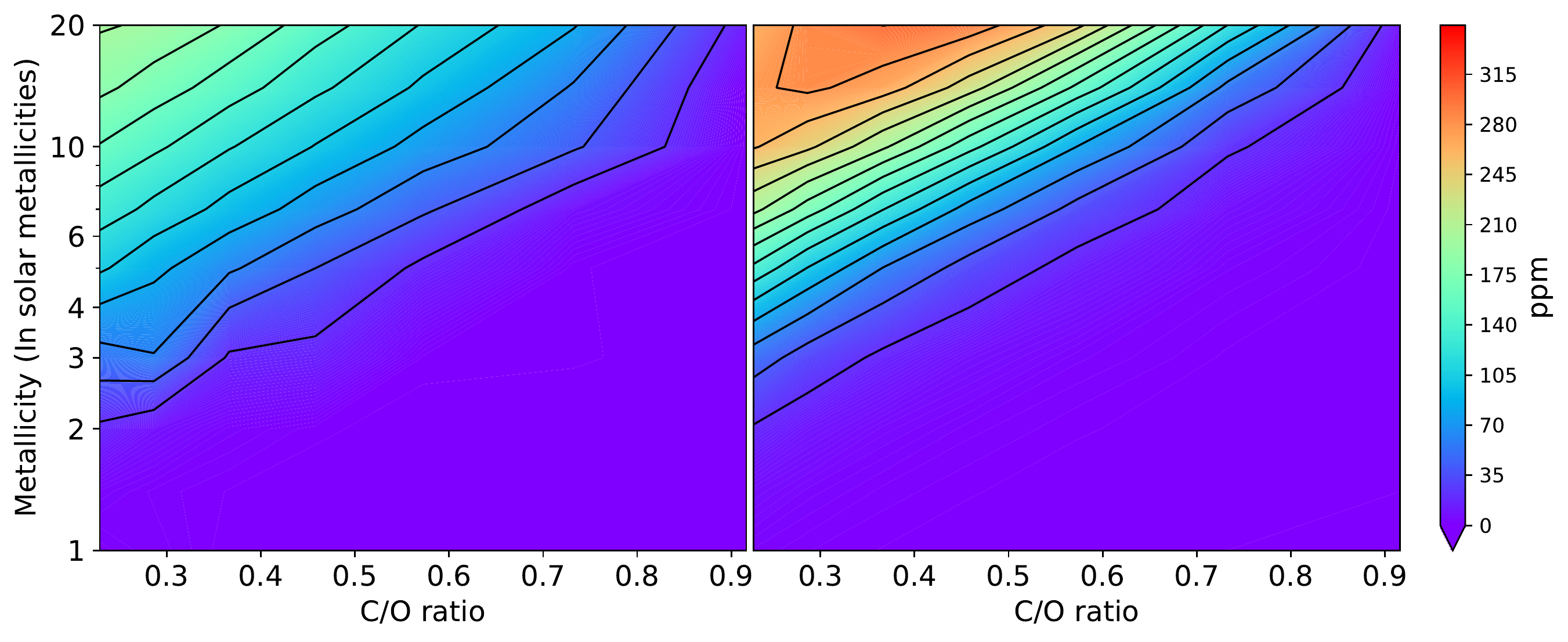}
        \caption{Contour plot of SO$_2$ detectability similar to Fig. \ref{SO2_Contour_Example} for two TP-profiles changed to be 200 K colder (\rw{left}) or hotter (\rw{right}) in order for all pressures. In both cases the C/O ratio is varied by changing the oxygen abundance.}
        \label{SO2_temp_change}
    \end{figure*}

For both H$_2$S and SO$_2$ we have only considered a single TP-profile each up to this point. To analyse the temperature dependence of the observed effects we did additional simulations on TP-profiles with atmospheres 200 K colder and hotter for all pressures. The results derived from this are shown in Figs. \ref{H2S_temp_change} and \ref{SO2_temp_change}. 

For H$_2$S we observe that the main structure is preserved in both cases. An increase in C/O ratio still leads to an increase in detectability until a metallicity dependent turning-point where the detectability falls off. Differences can be seen in the sharpness and overall height of the detectability peak. For the colder TP-profile the peak is a lot lower, peaking at 200 ppm and the peak is more spread overall to lower C/O ratios. For the hotter TP-profile the peak is a lot sharper, with a significantly lower detectability for lower C/O ratios, while the detectability peaks higher than for the original TP-profile, at 370 ppm. 

For SO$_2$ we observe that lowering the temperature leads to a significantly lower detectability. This is caused by an increase in the abundance of HCN. For this colder TP-profile the detectability peaks at 202 ppm. For the increased temperature we do see that the overall shape of the original is retained, although the values are slightly lower overall, leading to a maximum detectability of 300 ppm. For both species the overall characteristics seem to be retained within their 400 K window.

\subsection{Clouds}
\ym{The inclusion of clouds and hazes is out of the scope of this work. Nevertheless we note that} we do not consider clouds to be a problem for detecting SO$_2$ because of the low pressures at which it is abundant. Clouds at pressures above 10$^{-3}$ bar do not impact the detectability of SO$_2$, under the assumption that these clouds do not influence other atmospheric processes. For detection of H$_2$S complete cloud coverage at these pressures would cause a big problem, almost completely removing all effect of H$_2$S on the transmission spectrum. Complete cloud coverage above 10$^{-2}$ bar halves the detectability of H$_2$S on average, with complete cloud coverage above 10$^{-3}$ bar reducing detectability to less than 10 percent. To make definitive statements about the detectability of H$_2$S the likelihood of clouds needs to be considered for each case.

\subsection{The effect of the stellar flux}
In this paper we study the chemistry and detectability of sulfur-bearing species assuming as our fiducial case the planetary and host star parameters of HD 189733 b. Nevertheless, to consider the effect of the stellar flux on the photochemistry and detectability of H$_2$S and SO$_2$, we vary the orbital radius within VULCAN.
For H$_2$S, reducing the orbital radius by half, leads to a maximum loss of 5 percent in detectability for the relevant C/O ratios and metallicities with respect to the original analysis. The parametrized TP-profile is significantly hotter than expected for HD 189733 b, but this illustrates how the results vary little if the planet is located at a more reasonable distance for the used TP-profile. Halving or doubling the distance also only leads to a decrease of 5 percent in detectability at worst for SO$_2$ as well, although it can lead to small increases in certain situations as well. This result is quite interesting, since the creation of SO$_2$ is directly related to photodissociation. This change in orbital radius has very little impact on this. We find that the H$_2$O photodissociation and OH creation front shift only by a small amount, to higher or lower pressures, as a result of decreasing or increasing the orbital radius respectively. The increase in photodissociation from a decreased orbital radius provides more OH through the photodissociation of H$_2$O, forming SO$_2$ in larger amounts. This seems to be balanced out by the increased photodissociation of the SO$_2$ itself, although we have not been able to study this in detail.


We have also analysed the SO$_2$ detectability with regards to the stellar spectrum. In addition to our original spectrum for HD 189733 from \citet{Moses2011} we also test photospheric fits for HD 189733 and HD 209458, a significantly hotter star. To create an additional spectrum we take the difference between our original stellar spectrum and the photospheric fit for HD 189733, and add this difference to the photospheric fit of HD 209458. 

Compared to the original spectrum, detectability was found to be half in some cases when the spectrum of HD 209458 was used. Detectability decreased most when the original detectability was already low, while detectability only decreased by as little as 10 percent in cases where the original SO$_2$ detectability was high. A more detailed study of the impact of the stellar spectral shape is required to assess its full impact on SO$_2$ detectability.

\subsection{Eddy diffusion coefficient}
Within the model we have chosen for a constant value for the diffusion coefficient of 10$^9$ cm$^2$/s. This was done to simplify the situation within reasonable bounds. It is difficult to reach convergence for extremely small or large diffusion coefficients within VULCAN, but here we analyse the effect of decreasing the diffusion coefficient to 10$^8$ cm$^2$/s and of increasing the diffusion coefficient to 10$^{11}$ cm$^2$/s. Lowering the diffusion coefficient to 10$^8$ cm$^2$/s has little effect on the mixing ratios for models associated with either H$_2$S or SO$_2$ detection. This is in agreement with \citet{Hobbs2021}, which shows similar results for diffusion coefficients of $10^6$ and $10^9$ cm$^2$/s. For H$_2$S the detectability can decrease or increase by a few percent, with the overall detectability increasing slightly. For SO$_2$ the detectability increases for all situations, up to 10 percent. The effect of increasing the diffusion coefficient to 10$^{11}$ cm$^2$/s is significantly bigger. Many species become significantly more abundant at low pressure, including C$_2$H$_2$, HCN and H$_2$O, which can obscure H$_2$S, while the abundance of H$_2$S is barely affected at all. As a result the detectability of H$_2$S decreases up to 30 percent, although the decrease is smaller in most cases. 

\begin{figure*}
    \centering
    \includegraphics[width=17cm]{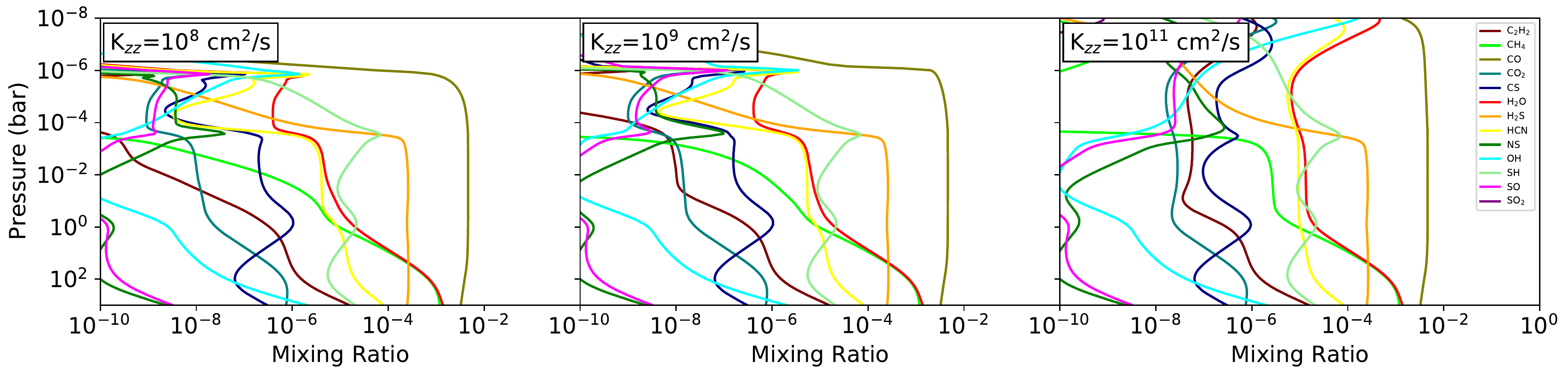}
        \caption{Comparison of the effect of a diffusion coefficient of 10$^8$, 10$^9$ and 10$^{11}$ cm$^2$/s on mixing ratios for 10 times solar metallicity, a C/O ratio of 0.96 and the TP-profile associated with H$_2$S detection.}
            \label{H2S_Kzz}
    \end{figure*}

\begin{figure*}
    \centering
    \includegraphics[width=17cm]{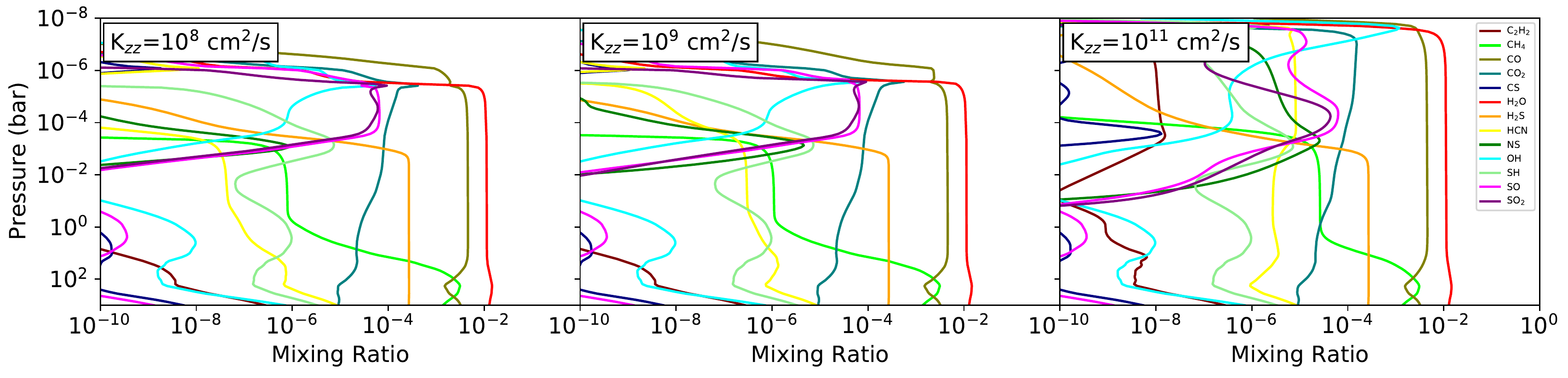}
        \caption{Comparison of the effect of a diffusion coefficient of 10$^8$, 10$^9$ and 10$^{11}$ cm$^2$/s on mixing ratios for 10 times solar metallicity, a C/O ratio of 0.29 and the TP-profile associated with SO$_2$ detection.}
            \label{SO2_Kzz}
    \end{figure*}

SO$_2$ is not obscured, both due to it being located higher in the atmosphere, and due it not being contested by other species at the relevant wavelengths. Instead for SO$_2$, its abundance is directly affected by the diffusion coefficient. The abundance of SO$_2$ drastically decreases for this higher diffusion coefficient. The SO/SO$_2$ ratio increases, with the abundance of SO being affected significantly less. \citet{VULCAN2021} similarly found that the effect of SO$_2$ on transmission spectra was significantly higher in the case of weak mixing compared to strong mixing for a model of GJ 436b. The decrease in SO$_2$ seems to be caused by photodissociation happening higher in the atmosphere, which results in OH being formed at lower pressures. At these pressures SO is significantly less abundant, making the formation of SO$_2$ less likely. This leads to decreases in detectability of up to 80 percent. The overall effect on the mixing ratios for H$_2$S and SO$_2$ are shown in Figs. \ref{H2S_Kzz} and \ref{SO2_Kzz} respectively.

\subsection{\ym{SO opacity}}
As mentioned before, we do not have opacity data for SO. While this limits our analysis by making it hard to predict anything regarding SO detection, we do not believe this has a significant negative impact on the detectability of other sulfur-bearing species. SO does not appear in large enough abundances to obscure H$_2$S in most scenarios, especially for the high C/O ratios where H$_2$S detectability is greatest. Including the opacity of SO could impact SO$_2$ detectability if it competes with it at the relevant wavelengths, . In that case the overall detectability of SO+SO$_2$ would be larger than that of SO$_2$ alone, but it would be difficult to make statements about their individual abundances. If SO does not compete with SO$_2$ at 7-8 $\mu$m it could still be observed at other wavelengths. It could also be obscured, but this is unlikely due to the low pressures at which we expect SO to be abundant. In this case any detection of SO will only aid the overall understanding of the content of hot Jupiter atmospheres and provide estimates of the SO/SO$_2$ ratio, which can also provide insight into SO$_2$ behaviour as mentioned previously. 

\subsection{\ym{Implications for planetary formation scenarios}}
Using our findings we can compare them to what we expect for C/O ratios and metallicities from planet formation models. Note that within these models metallicity refers to the abundance of elements heavier than helium, whereas until now we have used metallicity as a shorthand to refer to solar abundances, with the abundance of heavier elements scaled by a certain factor. SimAb \citep{SimAb2022} is a basic planet formation model incorporating gas and planetesimal accretion. The contents of gas and solids are dependent on the temperature at any given orbital radius and the input parameters of dust grain fraction and the \nk{planetesimal ratio} largely determine the resulting C/O ratios and metallicities. The study finds that C/O ratios of 0.8 can be reached in extreme cases with subsolar metallicities, while for supersolar metallicities the possible C/O ratios range \nk{from 0.2 to 0.65}. 

Putting the detection limit for a planet and central star similar to HD 189733 b at 100 ppm, this leaves a small range of possible C/O ratios and metallicities for detecting H$_2$S near a C/O ratio of 0.6 and solar metallicities. Since we do not have data for subsolar metallicities we cannot make predictions for H$_2$S detection in that region, even though the C/O ratio can be significantly higher in those cases. 

Using the same detection limit of 100 ppm for SO$_2$, the range of C/O ratios and metallicities for which SO$_2$ is detectable is significantly larger, with both C/O ratios below 0.4 with a metallicity of five times solar being detectable and C/O ratios until 0.5 reaching 100 ppm for metallicities above seven times solar. We have so far assumed that these metallicities, defined using the abundance of heavy elements, is equivalent to our previous definition of metallicity, using solar abundances, when analysing the detectability. In general this cannot be assumed, but since we also use the associated C/O ratio and sulfur is found to be overabundant compared to other elements at supersolar metallicities, we believe this to be a fair comparison. Additionally, putting the detection limit at 100 ppm is very strict, leaving room for variance as a result of the differences between abundances from the different definitions of metallicity.  

Another planet formation model \citep{Turrini2021} uses n-body simulations of growing and migrating planets in planetesimal disks to find elemental abundances for carbon, nitrogen, oxygen and sulfur for six formation scenarios differing by their initial core position. The C/O ratio varies very little from 0.49 to 0.58. The metallicity ranges from close to solar, for an initial core position of 5 AU, to roughly 8 times solar, for an initial core position of 130 AU. The relative abundances of carbon, oxygen and sulfur remain very close to solar, with only nitrogen becoming relatively less abundant for higher metallicities. As a result, we are again confident that we can compare this to our results, even though there is a discrepancy between the two definitions of metallicity. In none of the six formation scenarios do we predict to detect H$_2$S with the limit of 100 ppm. For detection of SO$_2$ the limit is only reached for scenario 6 with an initial core position of 130 AU. Both models do not incorporate atmospheric evolution, which could lead to a wider range of possible atmospheric compositions. 

\section{Conclusions}
In this work we analyse the detectability of sulfur-bearing species in hot Jupiter atmospheres. To achieve this we use the 1-D open-source photochemical kinetics code VULCAN to simulate such atmospheres. Low resolution transmission spectra are created using the modelling framework ARCiS based on the output of VULCAN. These results are placed into context by further analysing the dependence on temperature, cloud formation, the diffusion coefficient and planet formation models. 

We find that H$_2$S and SO$_2$ are the sulfur-bearing species most likely to be detected. H$_2$S is found in high abundances in most cases, but its detection relies on other species not being present in high enough abundances to obscure it. This happens at a temperature near 1500 K for C/O ratios between 0.7 and 0.9 depending on the metallicity. SO$_2$ can be formed in high abundances at \ym{temperatures close to 1000 K and low pressures}. Both low C/O ratios and high metallicities contribute to a high SO$_2$ abundance. Due to other species not showing large opacities near 7 $\mu$m and SO$_2$ only being found at low pressures, means that it is unlikely to be obscured.

Results for both H$_2$S and SO$_2$ are relatively stable for a temperature decrease or increase of 200 K. Results also stay consistent when decreasing the eddy diffusion coefficient. Increasing the diffusion coefficient to 10$^{11}$ cm$^2$/s shows a significant decrease in detectability for both H$_2$S and SO$_2$, although SO$_2$ is affected more. H$_2$S detectability can be affected by the presence of clouds, while SO$_2$ is not affected. H$_2$S is barely affected by a change in stellar flux, while SO$_2$ shows a significant effect. The shape of the stellar spectrum is largely irrelevant, with the overall luminosity being more important. Planet formation models indicate that it is unlikely for atmospheres to have the ideal C/O ratios and metallicities for H$_2$S detection, although detection could still be possible. SO$_2$ is most likely easier to detect, due to planet formation models favouring the formation of low C/O ratio, high metallicity planets.

\begin{acknowledgements}
We gratefully acknowledge support from the VULCAN development team. We thank the referee for useful comments and suggested additions to this paper. Y. Miguel and L.B.F.M. Waters are part of an international ISSI team.
\end{acknowledgements}

%
%

\bibliographystyle{aa}
\bibliography{bibliography.bib}

\begin{thebibliography}{20}
\expandafter\ifx\csname natexlab\endcsname\relax\def\natexlab#1{#1}\fi

\bibitem[{{Cazaux} {et~al.}(2022){Cazaux}, {Carrascosa}, {Mu{\~n}oz Caro},
  {Caselli}, {Fuente}, {Navarro-Almaida}, \&
  {Rivi{\'e}re-Marichalar}}]{cazaux2022}
{Cazaux}, S., {Carrascosa}, H., {Mu{\~n}oz Caro}, G.~M., {et~al.} 2022, \aap,
  657, A100

\bibitem[{{Cernicharo} {et~al.}(2021){Cernicharo}, {Cabezas}, {Ag{\'u}ndez},
  {Tercero}, {Pardo}, {Marcelino}, {Gallego}, {Tercero}, {L{\'o}pez-P{\'e}rez},
  \& {de Vicente}}]{cernicharo2021}
{Cernicharo}, J., {Cabezas}, C., {Ag{\'u}ndez}, M., {et~al.} 2021, \aap, 648,
  L3

\bibitem[{{Chubb} {et~al.}(2021){Chubb}, {Rocchetto}, {Yurchenko}, {Min},
  {Waldmann}, {Barstow}, {Molli{\`e}re}, {Al-Refaie}, {Phillips}, \&
  {Tennyson}}]{2021A&A...646A..21C}
{Chubb}, K.~L., {Rocchetto}, M., {Yurchenko}, S.~N., {et~al.} 2021, \aap, 646,
  A21

\bibitem[{{Codella} {et~al.}(2021){Codella}, {Bianchi}, {Podio}, {Mercimek},
  {Ceccarelli}, {L{\'o}pez-Sepulcre}, {Bachiller}, {Caselli}, {Sakai}, {Neri},
  {Fontani}, {Favre}, {Balucani}, {Lefloch}, {Viti}, \&
  {Yamamoto}}]{codella2021}
{Codella}, C., {Bianchi}, E., {Podio}, L., {et~al.} 2021, \aap, 654, A52

\bibitem[{Guillot(2010)}]{Guillot2010}
Guillot, T. 2010, Astronomy and Astrophysics, 520, A27

\bibitem[{Hobbs {et~al.}(2021)Hobbs, Rimmer, Shorttle, \&
  Madhusudhan}]{Hobbs2021}
Hobbs, R., Rimmer, P.~B., Shorttle, O., \& Madhusudhan, N. 2021, Monthly
  Notices of the Royal Astronomical Society, 506, 3186

\bibitem[{{Kama} {et~al.}(2019){Kama}, {Shorttle}, {Jermyn}, {Folsom},
  {Furuya}, {Bergin}, {Walsh}, \& {Keller}}]{kama2019}
{Kama}, M., {Shorttle}, O., {Jermyn}, A.~S., {et~al.} 2019, \apj, 885, 114

\bibitem[{{Khorshid} {et~al.}(2021){Khorshid}, {Min}, {D{\'e}sert}, {Woitke},
  \& {Dominik}}]{SimAb2022}
{Khorshid}, N., {Min}, M., {D{\'e}sert}, J.~M., {Woitke}, P., \& {Dominik}, C.
  2021, arXiv e-prints, arXiv:2111.00279

\bibitem[{{Laas} \& {Caselli}(2019)}]{laas2019}
{Laas}, J.~C. \& {Caselli}, P. 2019, \aap, 624, A108

\bibitem[{{Le Gal} {et~al.}(2021){Le Gal}, {{\"O}berg}, {Teague}, {Loomis},
  {Law}, {Walsh}, {Bergin}, {M{\'e}nard}, {Wilner}, {Andrews}, {Aikawa},
  {Booth}, {Cataldi}, {Bergner}, {Bosman}, {Cleeves}, {Czekala}, {Furuya},
  {Guzm{\'a}n}, {Huang}, {Ilee}, {Nomura}, {Qi}, {Schwarz}, {Tsukagoshi},
  {Yamato}, \& {Zhang}}]{legal2021}
{Le Gal}, R., {{\"O}berg}, K.~I., {Teague}, R., {et~al.} 2021, \apjs, 257, 12

\bibitem[{{Lodders} {et~al.}(2009){Lodders}, {Palme}, \& {Gail}}]{Lodders2009}
{Lodders}, K., {Palme}, H., \& {Gail}, H.~P. 2009, Landolt B\"ornstein, 4B, 712

\bibitem[{{Madhusudhan}(2019)}]{madhusudhan2019}
{Madhusudhan}, N. 2019, \araa, 57, 617

\bibitem[{Min {et~al.}(2020)Min, Ormel, Chubb, Helling, \&
  Kawashima}]{ARCiS2020}
Min, M., Ormel, C.~W., Chubb, K., Helling, C., \& Kawashima, Y. 2020, Astronomy
  and Astrophysics, 642, A28

\bibitem[{{Moses} {et~al.}(2011){Moses}, {Visscher}, {Fortney}, {Showman},
  {Lewis}, {Griffith}, {Klippenstein}, {Shabram}, {Friedson}, {Marley}, \&
  {Freedman}}]{Moses2011}
{Moses}, J.~I., {Visscher}, C., {Fortney}, J.~J., {et~al.} 2011, \apj, 737, 15

\bibitem[{Ormel \& Min(2019)}]{ARCiS2019}
Ormel, C.~W. \& Min, M. 2019, Astronomy and Astrophysics, 622, A121

\bibitem[{{Tsai} {et~al.}(2017){Tsai}, {Lyons}, {Grosheintz}, {Rimmer},
  {Kitzmann}, \& {Heng}}]{VULCAN2017}
{Tsai}, S.-M., {Lyons}, J.~R., {Grosheintz}, L., {et~al.} 2017, \apjs, 228, 20

\bibitem[{Tsai {et~al.}(2021)Tsai, Malik, Kitzmann, Lyons, Fateev, Lee, \&
  Heng}]{VULCAN2021}
Tsai, S.-M., Malik, M., Kitzmann, D., {et~al.} 2021, The Astrophysical Journal,
  923, 264

\bibitem[{Turrini {et~al.}(2021)Turrini, Schisano, Fonte, Molinari, Politi,
  Fedele, Pani{\'{c} }, Kama, Changeat, \& Tinetti}]{Turrini2021}
Turrini, D., Schisano, E., Fonte, S., {et~al.} 2021, The Astrophysical Journal,
  909, 40

\bibitem[{Wang {et~al.}(2017)Wang, Miguel, \& Lunine}]{Wang2017}
Wang, D., Miguel, Y., \& Lunine, J. 2017, The Astrophysical Journal, 850, 199

\bibitem[{Zahnle {et~al.}(2009)Zahnle, Marley, Freedman, Lodders, \&
  Fortney}]{Zahnle2009}
Zahnle, K., Marley, M.~S., Freedman, R.~S., Lodders, K., \& Fortney, J.~J.
  2009, The Astrophysical Journal, 701, L20

\end{thebibliography}


\end{document}